\newcommand{\beq}{\begin{equation}}
\newcommand{\eeq}{\end{equation}}
\newcommand{\bea}{\begin{eqnarray}}
\newcommand{\eea}{\end{eqnarray}}
\newcommand{\bear}{\begin{eqnarray*}}
\newcommand{\eear}{\end{eqnarray*}}

\newcommand{\rf}[1]{(\ref{#1})}
\documentstyle[aps,preprint,doublespace]{revtex}
\begin{document}

\draft

\title
{Exact Solution of Asymmetric Diffusion With $N$ Classes of 
 Particles of Arbitrary Size and
Hierarchical Order}

\author{
F. C. Alcaraz}  
\address{Departamento de F\'{\i}sica, 
Universidade Federal de S\~ao Carlos, 13565-905, S\~ao Carlos, SP
Brazil}
\author{ R. Z. Bariev}

\address{Departamento de F\'{\i}sica, 
Universidade Federal de S\~ao Carlos, 13565-905, S\~ao Carlos, SP
Brazil}

\address{The Kazan Physico-Technical Institute of the Russian 
Academy of Sciences, 
Kazan 420029, Russia}

\maketitle

\begin{abstract}

The exact solution of the asymmetric exclusion problem with  
$N$ distinct classes of particles ($c = 1,2,\dots,N$), with 
hierarchical order is presented. 
 In this model the particles 
(size 1)  are located at lattice points, and diffuse with  
equal asymmetric rates, but particles in a class $c$  do not distinguish 
those in the classes $c' >c$  from holes (empty sites). We generalize and 
solve exactly this model by considering the molecules in each  
 distinct class $c =1,2,\dots,N$ with sizes $s_c$  ($s_c = 0,1,2,\ldots$), in  
units of lattice spacing.	The solution is derived by 
a Bethe ansatz of nested type. 

\end{abstract}
\vspace{0.5cm}

\pacs{
}

\narrowtext    
\section{Introduction}

	The similarity between the master equation describing time 
fluctuations in nonequilibrium problems and the Schr\"odinger equation describing the 
quantum fluctuations of quantum spin chains turns out to be fruitful 
for both areas of research \cite{lushi}-\cite{alc-bar1}. 
Since many quantum chains are known to be 
exactly integrable through the Bethe ansatz, this provides exact information 
on the related stochastic model. At the same time the classical physical  
intuition and probabilistic methods successfully applied to nonequilibrium 
systems give new insights into the understanding of the physical and 
algebraic properties of quantum chains.

An example of this fruitful interchange is the problem of 
asymmetric diffusion of hard-core particles on the one 
dimensional lattice ( see \cite{der3,ligget,schu-domb} for reviews). 
This model 
is related to the exactly integrable anisotropic Heisenberg 
chain in its ferromagnetic regime \cite{yang} 
(XXZ model). However if we demand this quantum chain to be invariant 
under a quantum group symmetry $U_q(SU(2))$, we have to introduce, for the 
equilibrium statistical system, unusual surface terms, which on the other 
hand have a nice and simple interpretation for the related 
stochastic system \cite{alcrit1,alcrit2}.

In the area of exactly integrable models it is well known that one of the 
possible extensions of the spin-$\frac{1}{2}$ XXZ chain to higher spins, is 
the anisotropic spin-S Sutherland model (grading $\epsilon_1 = \epsilon_2 = 
\dots = \epsilon_{2s+1} = 1$) \cite{sutherland}. 
On the other hand in the area of diffusion limited 
reactions a simple extension of the asymmetric diffusion problem is the 
problem of diffusion with  particles belonging to $N$ distinct classes 
$(c =1,2,\dots,N)$ with hierarchical order 
\cite{bold}-\cite{ferra2} . In this 
problem a mixture  of hard-core particles 
diffuses on the lattice. Particles belonging to a class $c$ ($c =1,\dots,N$) 
 ignore 
the presence of those in classes $c' >c$, i. e., they see them in the 
same way as they see the holes (empty sites). In \cite{alcrit1} 
it was shown that for 
open boundary conditions the anisotropic spin-1 Sutherland model and this 
last stochastic model, in the case $N=2$, 
 are exactly related. The Hamiltonian governing 
the quantum or time fluctuations of 
both models being given in terms of generators 
of a Hecke algebra, invariant under the quantum group $U_qSU(3)$. 
In fact this relation can be extended to arbitrary values of $N$, and  
the quantum 
chain associated to the stochastic model is invariant under the quantum
$U_q(SU(N+1))$ group.
In this 
paper we are going to derive through the Bethe ansatz 
 the exact solution of the associated quantum 
chain, in a closed lattice. Recently \cite{alc-bar1} (see also \cite{sasawada})
we have shown that without losing 
its exact integrability, we can consider the problem of asymmetric 
diffusion with an arbitrary mixture of molecules 
with different sizes (even zero),  as 
long they do not interchange positions, that is, there is no reactions. In this 
paper  
 we are going to extend the asymmetric diffusion 
problem with $N$ type of particle with hierarchical order,  to the case where the particles in 
each class have an arbitrary size, in units of the lattice spacing. Unlike
the case of asymmetric diffusion problem, we have in this case a 
nested Bethe ansatz \cite{takh}. 
A pedagogical presentation for the simplest case $N=2$ was presented in 
\cite{revbra1}.

The paper is organized as follows. In the next section we introduce the 
generalized asymmetric model with $N$ type of particle with hierarchical order  
and derive the 
associated quantum chain. In section 3 the Bethe ansatz solution of the 
model is presented.
 Finally in section 4 
we present our conclusions, with some possible  generalizations of the 
stochastic problem considered in this paper, and some perspectives on
future work.

\section{The generalized asymmetric diffusion model
with $N$ classes of particles with hierarchical order}

	A simple extension of the asymmetric exclusion model, where hard-core 
particles diffuse on the lattice, is the problem where a mixture of 
particles belonging to different classes ($c =1,2,\dots,N$) diffuses 
on the lattice. This problem in the case where we have only $N=2$ classes 
was used to describe shocks 
\cite{bold}-\cite{ferra2}  in nonequilibrium 
and also has a stationary probability distribution that can be expressed  
via the matrix-product ansatz \cite{derrida1}. 
In \cite{mallick} it was also shown that the stationary state 
of the case $N=3$ can also be expressed by the matrix-product ansatz. 
In this model we have $n_1, n_2, \dots n_N$  
molecules belonging to the classes $c=1,2,\dots,N$, respectively. All classes 
of molecules diffuse asymmetrically, but with the same asymmetrical rates, 
whenever they encounter empty sites (holes) at nearest-neighbor sites. However,
when molecules of different classes, $c$ and $c'$ ($c <c'$),  are at their 
minimum separation, the  
molecules of class $c$  exchange position with the same rate as they 
diffuse, and consequently the  molecules in the class $c$ see no difference 
between molecules belonging to the classes $c'>c$  and holes. 

We now introduce a generalization of the above model, where instead of having
unit size, the molecules in each distinct class $c=1,2,\dots,N$  
have in general distinct sizes $s_1,s_2,\dots,s_N$   
($s_1,\dots,s_N =1,2,\ldots$), 
respectively,
in units of lattice spacing. In Fig. 1
we show some examples of molecules of different sizes. We may think of a  
molecule of size $s$ as  formed by $s$ monomers (size 1), and for 
simplicity, we define the position of the molecule as the center 
of its leftmost monomer. The molecules have a hard-core repulsion: the 
minimum distance $d_{\alpha \beta}$, in units of the lattice spacing, between 
molecules $\alpha$ and $\beta$, with $\alpha$ in the left, is given by 
$d_{\alpha\beta} =s_{\alpha}$.  In order to describe the occupancy of a given 
configuration of molecules we attach at each lattice site $i$ ($i =1,2,
\dots,L$) a variable 
 $\beta_i$ ($i=1,2,\ldots,L$), taking the values 
$\beta_i = 0,1,\dots,N$. 
The values $\beta=1,2,\ldots,N$ represent the sites occupied by 
molecules of class $c=1,2,\ldots,N$, respectively. 
On the other hand the value $\beta =0$ represent the empty 
sites as well the excluded ones, due to the finite size of the 
molecules. As an example, in a $L=8$ sites chain, the configuration 
where a particle of class 1, with size $s_1=2$ located at 
site 1, and another particle of class 2, with size $s_2 =3$ 
located at site $3$, is represented by 
$\{\beta\} = \{1,0,0,2,0,0,0,0\}$.
 Then the allowed configurations are given by the set 
$\{\beta_i\}$ ($i=1,\ldots,L$), where for each pair 
$(\beta_i,\beta_j)\ne 0$ with $j>i$ we should have $j -i \geq s_{\beta_i}$.

The time evolution of the probability distribution 
$ P(\lbrace\beta\rbrace,t)$, 
of a given configuration $\{\beta\}$ is given by the master equation 
\beq
\label{1} 
\frac{\partial  P(\lbrace\beta\rbrace,t)}{\partial t} =
\sum_{\{\beta'\}}\left[ 
- \Gamma(\lbrace\beta\rbrace 
\rightarrow \lbrace\beta'\rbrace)  P(\lbrace\beta\rbrace,t) +
\Gamma(\lbrace\beta'\rbrace 
\rightarrow \lbrace\beta\rbrace)  P(\lbrace\beta'\rbrace,t),
\right]
\eeq
aqui
where $\Gamma(\lbrace\beta\rbrace \rightarrow \lbrace\beta'\rbrace)$ is 
the transition rate for configuration $\lbrace\beta\rbrace$ to change to
$\lbrace\beta'\rbrace$.  In the present model we only allow,
whenever 
the constraint of excluded volume is satisfied,  
the particles  to diffuse to  
nearest-neighbor sites, or to exchange positions. 
The possible motions are diffusion to the right
\beq
\label{2}
\beta_i\;\;\emptyset_{i+1} \rightarrow \emptyset_i\;\;\beta_{i+1},
\quad (\beta =1,\dots,N) \;\;\;\;\;\;\;\;
\;\;\;\;\;\;\;(\mbox {rate} \;\; \Gamma_R)
\eeq
diffusion to the left 
\beq
\label{3}
\emptyset_i\;\;\beta_{i+1} \rightarrow \beta_i\;\;\emptyset_{i+1},
\quad (\beta =1,\dots,N) \;\;\;\;\;\;\;\;
\;\;\;\;\;\;\;(\mbox {rate} \;\; \Gamma_L)
\eeq
and interchange of particles
\bea 
\label{4}
\beta_i\;\;\beta'_{i+s_{\beta}} &\rightarrow& \beta'_i\;\;\beta_{i+s_{\beta'}},
\quad (\beta <\beta' =1,\dots,N) \;\;\;\;\;\;\;\;
\;\;\;\;\;\;\;(\mbox {rate}\;\;  \Gamma_R)\nonumber\\
\beta_i\;\;\beta'_{i+s_{\beta}} &\rightarrow& \beta'_i\;\;\beta_{i+s_{\beta'}},
\quad (\beta >\beta' =1,\dots,N) \;\;\;\;\;\;\;\;
\;\;\;\;\;\;\;(\mbox {rate}\;\;  \Gamma_L).
\eea
As we see from \rf{4}, particles 
belonging to a given class $c$ interchange positions with those of 
class $c'>c$ 
 with the same rate as they interchange positions with 
the empty sites (diffusion). We should remark however that unless the  
  particles in class $c'$ have unit size ($s_c =1$), the net effect of these 
 particles in those of class $c$ is distinct from the effect 
produced by the holes, since as the result of the exchange the particles in
class $c$ 
 will move by $s_{c'}$ lattice size units, accelerating its diffusion.

 The master equation \rf{1} can be written as 
a Schr\"odinger equation in Euclidean time (see Ref. \cite{alcrit1} for general
application for two body processes)
\beq
\label{5}
\frac{\partial| P>}{\partial t} = -H | P>,
\eeq
if we interpret $| P> \equiv  P(\lbrace\beta\rbrace, t)$ 
as the associated wave 
function. If we represent $\beta_i$ as $|\beta>_i$ the vector 
$|\beta>_1\otimes |\beta>_2 \otimes \cdots \otimes |\beta>_L$ will give us 
the associated 
Hilbert space. The process \rf{2}-\rf{4} gives us the Hamiltonian (see 
Ref. \cite{alcrit1} for general applications)
\bea
\label{6}
H &=& D \sum_{j} H_{j}  \nonumber \\
H_j &=& - {\cal P} \{ \sum_{\alpha=1}^N \left[ \epsilon_+(E_{j}^{0\alpha}
E_{j+1}^{\alpha 0} - E_j^{\alpha \alpha} E_{j+1}^{0 0}) +
\epsilon_-(E_j^{\alpha 0}E_{j+1}^{0\alpha} - E_j^{0 0} E_{j+1}^{\alpha \alpha}) 
\right] \nonumber \\
&& +\sum_{\alpha =1}^N \sum_{\beta=1}^N \epsilon_{\alpha,\beta}
(E_j^{\beta \alpha}E_{j+s_{\beta}}^{\alpha 0} E_{j +s_{\alpha}}^{0 \beta} 
-E_j^{\alpha \alpha} E_{j +s_{\beta}}^{0 0} E_{j+s_{\alpha}}^{\beta \beta}
)\}{\cal P}
\eea
with 
\beq
\label{7}
D = \Gamma_R + \Gamma_L,\;\;\; \epsilon_+ =\frac{\Gamma_R}{\Gamma_R + 
\Gamma_L},\;\;\; \epsilon_- = \frac{\Gamma_L}{\Gamma_R + \Gamma_L} \quad 
(\epsilon_+ + \epsilon_- = 1),
\eeq
\beq
\label{7a}
\epsilon_{\alpha \beta} =\cases{\epsilon_+ & $\alpha < \beta$\cr 
0 & $\alpha = \beta$ \cr
\epsilon_- & $\alpha >\beta$ \cr} 
\eeq
and periodic boundary conditions. 
The matrices $E^{\alpha,\beta}$ are $(N+1)\times 
(N+1)$ matrices with a single nonzero
element $(E^{\alpha,\beta})_{i,j} = \delta_{\alpha,i}\delta_{\beta,j}$  
($\alpha,\beta,i,j=0,\dots,N)$. The projector ${\cal P}$ in \rf{6}, 
projects out from the associated  Hilbert 
space the vectors  $|\lbrace\beta\rbrace>$ which represent forbidden 
positions of the molecules due to their finite size, which mathematically 
 means that for all $i,j$ with
$\beta_i,\beta_j \ne 0,\;\; |i - j| \ge  s_{\beta_i} \;\;(j >i)$.
The constant $D$ in  \rf{6} fixes the time scale and for simplicity we
 chose $D = 1$. A particular simplification of \rf{6}
occurs when  the molecules in all classes  have the same  size  
$s_1 =s_2 =\dots =s_N = s$. In this case
the Hamiltonian can be expressed as an anisotropic nearest-neighbor 
interaction spin-$N/2$ $SU(N+1)$ chain. Moreover in the case where their  
sizes are unity  ($s=1$) the model  can be related to the 
anisotropic version \cite{tsveli} of the $SU(N+1)$ 
Sutherland 
model \cite{sutherland} with twisted boundary conditions.

\section{ The Bethe ansatz equations}

We present in this section the exact solution of the general quantum chain 
\rf{6}. 
A pedagogical presentation for the particular case where $N=2$  
was presented in \cite{revbra1}.

Due to the conservation of particles in the diffusion and interchange 
processes the 
total number of particles $n_1,n_2,\dots,n_N$  in each class  
are good quantum numbers and consequently we can split 
the associated Hilbert space into block disjoint sectors labeled by the 
numbers $n_1,n_2\dots n_N$ ($n_i = 0,1,\ldots; i =1,\dots,N$).
We therefore consider the eigenvalue equation 
\beq
\label{8}
H |n_1,n_2,\dots,n_N> = E|n_1,n_2\dots,n_N>,
\eeq
where
\beq
\label{9}
|n_1,n_2,\dots,n_N> = \sum_{\{Q\}} \sum_{\{x\}}f(x_1,Q_1;\ldots;x_n,Q_n)
|x_1,Q_1;\ldots;x_n,Q_n>,
\eeq
and $n = \sum_{i=1}^N n_i$ is the total number of particles. In \rf{9} 
 $|x_1,Q_1;\ldots;x_n,Q_n>$ means the configuration where a 
particle of class $Q_i$ ($Q_i=1,2,\dots,N$) 
is at position $x_i$ ($x_i =1,\ldots,L$). The summation $\{Q\} = \{Q_1,\ldots,
Q_n\}$ extends over all permutations of the $n$ integer numbers
$\{1,2,\dots,N\}$  
in which $n_i$ terms have the value $i$ ($i=1,2,\dots,N$) 
, while the summation $\{x\} =\{x_1,\ldots,x_n\}$ runs, for each 
permutation $\{Q\}$, in the set of 
the $n$ nondecreasing 
integers satisfying 
\beq
\label{10}
x_{i+1} \geq x_{i} + s_{Q_i}, \quad i=1,\ldots,n-1, \quad s_{Q_1} 
\leq x_n -x_1 \leq N-s_{Q_n}.
\eeq
Before getting the results for general values of $n$ let us consider initially 
the cases where we have 1 or 2 particles.

{\it {\bf  n = 1.}}
For one particle on the chain, in any class $c=1,2,\dots,N$, 
as a consequence of the translational invariance 
of \rf{6} it is simple to verify directly that 
the eigenfunctions are the momentum-$k$ eigenfunctions 
\beq
\label{11}
|0,\dots,0,1_c,0,\dots,0> = 
\sum_{x=1}^{L} f(x,c) |x,c>, \quad c =1,\dots,N 
\eeq
with
\beq
\label{12}
 f(x,c) =   e^{ikx},\quad k= \frac{2\pi l}{L}, \quad 
l = 0,1,\ldots,L-1,
\eeq
and energy given by 
\beq
\label{13}
E = e(k) \equiv  -(\epsilon_-e^{ik} + \epsilon_+e^{-ik} -1).
\eeq

{\it {\bf  n =2.}} For two particles of classes $Q_1$ and 
$Q_2$ ($Q_1,Q_2=1,2,\dots,N$) on the lattice, the eigenvalue equation \rf{8} 
 gives  us two distinct relations depending on the relative location of 
the particles. The first relation applies to the case in which a 
particle of class $Q_1$ (size $s_{Q_1}$) is at position $x_1$ and a 
particle $Q_2$ (size $s_{Q_2}$) is at position $x_2$, where 
 $x_2 >x_1 +s_{Q_1}$. We obtain in this case the relation 
\bea
\label{14}
Ef(x_1,Q_1;x_2,Q_2) &=& -\epsilon_+f(x_1-1,Q_1;x_2,Q_2) - 
\epsilon_-f(x_1,Q_1;x_2+1,Q_2) \nonumber \\ 
-\epsilon_-f(x_1+1,Q_1;x_2,Q_2) 
 &-&\epsilon_+f(x_1,Q_1;x_2-1,Q_2) + 2f(x_1,Q_1;x_2,Q_2),
\eea
where we have used the relation $\epsilon_+ + \epsilon_- = 1$. This last 
equation
 can be solved promptly by the ansatz
\bea
\label{17}
f(x_1,Q_1;x_2,Q_2) &=& \sum_P A_{P_1,P_2}^{Q_1,Q_2}e^{i(k_{P_1}x_1 + 
k_{P_2}x_2)}   \nonumber \\
&=& A_{1,2}^{Q_1,Q_2}e^{i(k_1x_1+k_2x_2)} + 
A_{2,1}^{Q_1,Q_2}e^{i(k_2x_1+k_1x_2)}
\eea
with energy
\beq
\label{16}
E = e(k_1) + e(k_2),
\eeq
where $k_1,k_2,A_{1,2}^{Q_1,Q_2}$ and $A_{2,1}^{Q_1,Q_2}$ are free 
parameters to be fixed.
In \rf{17} the summation is over the  
permutations  $P=P_1,P_2$ of  
(1,2). The second relation applies when 
$x_2 = x_1 + s_{Q_1}$. In this case instead of \rf{14} we have 
\bea
\label{18}
Ef(x_1,Q_1;&&x_1 +s_{Q_1},Q_2) = -\epsilon_+f(x_1-1,Q_1;x_1+s_{Q_2},Q_2) - 
\epsilon_-f(x_1,Q_1;x_1+s_{Q_1}+1,Q_2)
 \nonumber \\
&&- \tilde \epsilon_{Q_2,Q_1}f(x_1,Q_2;x_1 +s_{Q_2},Q_1) + 
(1 + \tilde \epsilon_{Q_1,Q_2})f(x_1,Q_1;x_1 +s_{Q_1},Q_2).
\eea
If we now substitute the ansatz \rf{17} with the energy \rf{16}, the 
constants $A_{12}^{Q_1,Q_2}$ and $A_{21}^{Q_1,Q_2}$, initially arbitrary, 
should now satisfy 
\beq
\label{20}
\sum_P \{ \left[ {\cal D}_{P_1,P_2} + e^{ik_{P_2}}(1- \tilde
\epsilon_{Q_1,Q_2})\right]e^{ik_{P_2}(s_{Q_1}-1)}A_{P_1,P_2}^{Q_1,Q_2} +  
\tilde \epsilon_{Q_2,Q_1} e^{ik_{P_2}s_{Q_2}}A_{P_1,P_2}^{Q_2,Q_1}\} = 0
\eeq
where

\beq
\label{21} 
{\cal D}_{l,m} = -(\epsilon_+ + \epsilon_-e^{i(k_l +k_m)}).
\eeq
At this point it is convenient to consider separately the case where $Q_1=Q_2$ 
from those where $Q_1 \neq Q_2$. If $Q_1=Q_2=Q$ ($Q=1,\dots,N$) eq. \rf{20} gives
\beq
\label{22}
\sum_P \left(  {\cal D}_{P_1,P_2} + e^{ik_{P_2}} 
\right)e^{ik_{P_2}(s_{Q}-1)}A_{P_1,P_2}^{Q,Q} = 0
\eeq
and the cases $Q_1 \neq Q_2$ give us the equations
\bea 
\label{23}
\sum_P\left[ \begin{array}{cc} {\cal D}_{P_1,P_2}
+e^{ik_{P_2}}\epsilon_{Q_2,Q_1} &
\epsilon_{Q_2,Q_1}e^{ik_{P_2}}  \nonumber \\
\epsilon_{Q_1,Q_2}e^{ik_{P_2}} & {\cal D}_{P_1,P_2}+\epsilon_{Q_1,Q_2}e^{ik_{P_2}} \nonumber 
\end{array} \right]
\left[ \begin{array}{c} e^{ik_{P_2}(s_{Q_1}-1)}A_{P_1,P_2}^{Q_1,Q_2} \nonumber \\
e^{ik_{P_2}(s_{Q_2}-1)}A_{P_1,P_2}^{Q_2,Q_1} \end{array} \right] = 0.
\eea  
Performing the above summation we obtain, after lengthy but straightforward 
algebra, the following relation among the amplitudes 
\bea  \label{24}
\left[ \begin{array}{c} A_{1,2}^{Q_1,Q_2}e^{ik_2(s_{Q_1}-1)} \nonumber \\
A_{1,2}^{Q_2,Q_1}e^{ik_2(s_{Q_2}-1)} \end{array} \right] &=& 
-\frac{{\cal D}_{1,2} + e^{ik_1}} {{\cal D}_{1,2} +e^{ik_2}}  \nonumber 
\\ &\times & 
\left\{ 1 - \Phi(k_1,k_2)\left[ \begin{array} {cc} \epsilon_{Q_1,Q_2} 
& -\epsilon_{Q_2,Q_1}
\nonumber \\
 -\epsilon_{Q_1,Q_2} & \epsilon_{Q_2,Q_1} \end{array} \right] \right\}
\left[ \begin{array} {c} 
A_{2,1}^{Q_1,Q_2}e^{ik_1(s_{Q_1}-1)} \nonumber \\
A_{2,1}^{Q_2,Q_1}e^{ik_1(s_{Q_2}-1)} \nonumber \end{array} \right],
\eea
where
\beq \label{25}
\Phi(k_1,k_2) = \frac{e^{ik_1} - e^{ik_2}}{{\cal D}_{1,2} +e^{ik_1}} .
\eeq
Equations \rf{22} and \rf{24} can be written in a compact form 
\beq \label{26}
A_{P_1,P_2}^{Q1,Q2} = -\Xi_{P_1,P_2} \sum_{Q_1',Q_2' =1}^N 
S_{Q_1',Q_2'}^{Q_1,Q_2}(k_{P_1},k_{P_2})A_{P_2,P_1}^{Q_2',Q_1'}, \quad 
(Q_1,Q_2 =1,\dots,N)
\eeq
with
\beq \label{27}
\Xi_{l,j} = \frac{{\cal D}_{l,j} + e^{ik_l}} {{\cal D}_{l,j} +e^{ik_j}} =
\frac {\epsilon_+ + \epsilon_-e^{i(k_l+k_j)} -e^{ik_l}} 
{\epsilon_+ + \epsilon_-e^{i(k_l+k_j)} - e^{ik_j}},
\eeq
where we have introduced the $S$ matrix. From \rf{22} and \rf{24} this S matrix 
has only $N(2N-1)$  
non zero elements, namely 
\bea \label{28}
S_{Q_2,Q_1}^{Q_1,Q_2}(k_1,k_2)
 &=& \left[ 1-\epsilon_{Q_1,Q_2}\Phi(k_1,k_2) \right] e^{i(k_1-k_2)(s_{Q_1}-1)} 
\quad (Q_1,Q_2 = 1,\dots,N),
\nonumber \\
S_{Q_1,Q_2}^{Q_1,Q_2}(k_1,k_2) &=& 
\epsilon_{Q_2,Q_1}\Phi(k_1,k_2)  e^{ik_1(s_{Q_2} -1)} e^{-ik_2(s_{Q_1}-1)}
\quad (Q_1,Q_2=1,\dots,N; Q_1 \neq Q_2).
\eea
 Equations 
\rf{26} do not fix the ``wave numbers" $k_1$ and $k_2$. In general, these numbers 
are complex, and are fixed due to the cyclic boundary condition 
\beq
\label{29}
f(x_1,Q_1;x_2,Q_2) =  f(x_2,Q_2;x_1+N,Q_1),
\eeq
which from \rf{17} give the relations
\beq
\label{30}
A_{1,2}^{Q_1Q_2}  = e^{ik_1N}A_{2,1}^{Q_2,Q_1}, \quad \quad A_{2,1}^{Q_1,Q_2} = 
e^{ik_2N}A_{2,1}^{Q_2,Q_1}.
\eeq
This last equation, when solved by exploiting \rf{26}-\rf{28}, gives us the 
possible values of $k_1$ and $k_2$, and from \rf{16} the eigenenergies 
in the sector with 2 particles. Instead of solving these equations for the 
particular case $n=2$ let us now consider the case of general $n$.

{\it {\bf General n.}} The above calculation can be generalized for 
arbitrary 
occupation \\ 
$\{n_1,n_2,\dots,n_N\}$ of particles in classes $1,2,\dots,N$, respectively. 
  The ansatz for the wave function \rf{9}  becomes 
\beq
\label{31}
f(x_1,Q_1;\ldots;x_n,Q_n) = \sum_P A_{P_1,\ldots,P_n}^{Q_1,\cdots, Q_n}
 e^{i(k_{P_1}x_1+\cdots + 
k_{P_n}x_n)},
\eeq
where the sum extends over all permutations $P$ of the integers 
$1,2,\ldots,n$, and $n = \sum_{i=1}^N n_i$ is the total number 
of particles.

The application of the translation operator in the above wave functions
 implies that \rf{9} are also eigenfunctions of the momentum operator 
with eigenvalues 
\beq \label{o1}
p = \sum_{j=1}^n k_j = \frac{2\pi l}{L}, \quad (l=0,1,\dots,L-1).
\eeq
For the components $|x_1,Q_1;\ldots;x_n,Q_n>$ where  
$x_{i+1} -x_i >s_{Q_i}$ for $i=1,2,\ldots,n$, it is simple to see
that the eigenvalue equation \rf{8} is satisfied by the ansatz \rf{31} 
with energy 
\beq
\label{32}
E = \sum_{j=1}^n e(k_j).
\eeq
On the other hand if
 a pair of particles of class $Q_i,Q_{i+1}$ is at positions 
$x_i,\; x_{i+1}$, where $x_{i+1} = 
x_i +s_{Q_i}$, equation \rf{8} with the ansatz \rf{31} and the relation 
\rf{32} give us the generalization of relation \rf{26}, namely
\beq
\label{33}
A_{\ldots,P_i,P_{i+1},\ldots}^{\cdots, Q_i,Q_{i+1},\cdots} = 
 -\Xi_{P_i,P_{i+1}} \sum_{Q_1',Q_2'}^N S_{Q_1',Q_2'}^{Q_i,Q_{i+1}}
(k_{P_i},k_{P_{i+1}}) A_{\ldots, P_{i+1},P_i,\ldots}^{\cdots, Q_2',Q_1',\cdots} 
\quad \;\; (Q_i,Q_{i+1} =1,2,,\dots,N),
\eeq
with $S$ given by eq. \rf{28}. Inserting the ansatz \rf{31} in the 
boundary condition 
\beq
\label{34}
f(x_1,Q_1;\ldots;x_n,Q_n) = f(x_2,Q_2;\ldots;x_n,Q_n;x_1+N,Q_1)
\eeq
we obtain  the additional relation 
\beq
\label{35}
A_{P_1,\ldots,P_n}^{Q_1,\cdots, Q_n} = e^{ik_{P_1}N}
A_{P_2,\ldots,P_n,P_1}^{Q_2,\cdots, Q_n,Q_1},
\eeq
which together with \rf{33} should give us the energies.

 Successive applications of \rf{33} give us in 
general distinct relations between the amplitudes. For example $A_{\ldots,
k_1,k_2,k_3,\ldots}^{\ldots,\alpha,\beta,\gamma,\ldots}$ relate to 
$A_{\ldots,k_3,k_2,k_1,\ldots}^{\ldots,\gamma,\beta,\alpha,\ldots}$ by 
performing the permutations $\alpha\beta\gamma \rightarrow \beta\alpha\gamma 
\rightarrow \beta\gamma\alpha \rightarrow \gamma\beta\alpha$ or 
 $\alpha\beta\gamma \rightarrow \alpha\gamma\beta  
\rightarrow \gamma\alpha\beta \rightarrow \gamma\beta\alpha$, and consequently 
the $S$-matrix should satisfy the Yang-Baxter \cite{yang,baxter} equation
\bea
\label{36}
\sum_{\gamma,\gamma',\gamma''=1}^N S_{\gamma,\gamma'}^{\alpha,\alpha'}(k_1,k_2)
S_{\beta,\gamma''}^{\gamma,\alpha''}(k_1,k_3)& &S_{\beta',\beta''}^
{\gamma',\gamma''}(k_2,k_3) = \nonumber \\
& & \sum_{\gamma,\gamma',\gamma''=1}^N
 S_{\gamma',\gamma''}^{\alpha',\alpha''}(k_2,k_3)
S_{\gamma,\beta''}^{\alpha,\gamma''}(k_1,k_3)S_{\beta,\beta'}^
{\gamma,\gamma'}(k_1,k_2), 
\eea
for $\alpha,\alpha',\alpha'',\beta,\beta',\beta'' =1,2,\dots,N$ and 
 $S$ given by \rf{28}. 
  Actually the relation \rf{36} 
is a necessary and sufficient condition \cite{yang,baxter} 
to obtain a non-trivial 
solution for the amplitudes in Eq. \rf{33}. 

We can verify by a long and straightforward calculation that
for arbitrary number of 
classes $N$ and values of the sizes  $s_1,s_2,\dots,s_N$,
the $S$ matrix \rf{28}, satisfies the Yang-Baxter 
equation \rf{36}, and consequently we may use relations \rf{33} and \rf{35} 
to obtain the eigenenergies of the Hamiltonian \rf{6}. Applying relation  
\rf{33} $n$ times on the right of equation \rf{35} we obtain a relation 
between the amplitudes with the same ordering in the lower indices: 
\bea \label{37}
A_{P_1,\ldots,P_n}^{Q_1,\ldots,Q_n} = e^{ik_{P_1}N} A_{P_2,\ldots,P_n,P_1}^
{Q_2,\ldots,Q_n,Q_1} = \left(\prod_{i=2}^n -\Xi_{P_i,P_1}\right) 
e^{ik_{P_1}N} \sum_{Q_1',\ldots,Q_n'}\sum_{Q_1'',\ldots,Q_n''} \nonumber \\
S_{Q_1',Q_1''}^{Q_1,Q_2''}(k_{P_1},k_{P_1})
S_{Q_2',Q_2''}^{Q_2,Q_3''}(k_{P_2},k_{P_1})\cdots
S_{Q_{n-1}',Q_{n-1}''}^{Q_{n-1},Q_n''}(k_{P_{n-1}},k_{P_1})
S_{Q_{n}',Q_{n}''}^{Q_{n},Q_1''}(k_{P_{n}},k_{P_1})
A_{P_1,\ldots,P_n}^{Q_1',\ldots,Q_n'},
\eea
where we have introduced the harmless extra sum
\beq
\label{38}
1 = \sum_{Q_1'',Q_2''=1}^N \delta_{Q_2'',Q_1'} \delta_{Q_1'',Q_1} =
 \sum_{Q_1'',Q_2''=1}^N S_{Q_1',Q_1''}^{Q_1,Q_2''}(k_{P_1},k_{P_1}) 
\eeq
(see \cite{revbra1} for ilustrations of the above equations).
In order to fix the values of $\{k_j\}$ we should solve \rf{37}, i.e., 
we should find the eigenvalues $\Lambda(k)$ of the matrix
\beq
\label{39}
T(k)_{\{Q'\}}^{\{Q\}} = \sum_{Q_1'',\ldots,Q_n''=1}^N  \left(
\prod_{l=1}^{n} 
S_{Q_l',Q_l''}^{Q_l,Q_{l+1}''}(k_{P_l} ,k)\right) ,
\eeq
with periodic boundary condition
\beq \label{37p}
S_{Q'_n,Q''_n}^{Q_n,Q''_{n+1}}(k_{P_n},k) = 
S_{Q'_n,Q''_n}^{Q_n,Q''_{1}}(k_{P_n},k) .
\eeq 
The Bethe-ansatz equations which fix the set $\{k_l\}$ will be given from
 \rf{37} by 
\beq \label{40}
e^{-ik_jN} = (-1)^{n-1}\left( 
\prod_{l=1}^n\Xi_{l,j}\right)\Lambda(k_j), \quad j=1,\ldots,n.
\eeq
The matrix $T(k)$ has dimension $N^n \times N^n$ and can be interpreted 
as the transfer matrix of a inhomogeneous $N(2N-1)$-vertex model in a 
two dimensional lattice with periodic boundary condition in the 
horizontal direction ($n$ sites).
Due to the special form of the S matrix 
\rf{28} the eigenvalues of \rf{39} are invariant under a 
local gauge transformation where for each factor 
$S(k_{P_l},k)$ in \rf{39}:
\beq \label {b1}
S_{Q'_l,Q''_l}^{Q_l,Q''_{l+1}}(k_{P_l},k) \rightarrow 
S_{Q'_l,Q''_l}^{Q_l,Q''_{l+1}}(k_{P_l},k) \frac{\phi_{Q''_{l+1}}^{(l)}}
{\phi_{Q''_l}^{(l)}}, 
\eeq
where $\phi_{\alpha}^{(l)}$ ($l=1,\dots,L; \quad \alpha=1,\dots,N$) are 
arbitrary functions. If we perform the transformation \rf{b1} with 
the special choice
\beq \label{b11}
\frac{\phi_{\alpha}^{(l+1)}}{\phi_{\alpha}^{(l)}} = 
e^{-ik_{P_l}(s_{\alpha}-1)}, \quad (l=2,3,\dots,N),
\eeq
the equivalent transfer matrix to be diagonalized is given by 
\beq \label{b2}
\tilde T(k)_{\{Q'\}}^{\{Q\}} = e^{-ik\sum_{i=1}^n(s_{Q_i}-1) }
T_0(k)^{\{Q\}}_{\{Q'\}}
\eeq
where
\beq \label{b2a}
T_0(k)^{\{Q\}}_{\{Q'\}}= 
\sum_{Q_1'',\ldots,Q_n''=1}^N  \left(
\prod_{l=1}^{n} 
\tilde S_{Q_l',Q_l''}^{Q_l,Q_{l+1}''}(k_{P_l},k)\right) ,
\eeq
with the twisted boundary condition
\beq \label{b3}
\tilde S_{Q'_n,Q''_n}^{Q_n,Q''_{n+1}}(k_{P_n},k) = 
\tilde S_{Q'_n,Q''_n}^{Q_n,Q''_1}(k_{P_n},k) 
\Phi_{Q''_1}
\eeq
with twisted phase
\beq \label{b3a}
\Phi_l = 
e^{i(s_l-1)\sum_{j=1}^n k_j}, \, \, l=1,\ldots,N.
\eeq
The matrix $\tilde S$  in \rf{b2a} and \rf{b3} 
is obtained from those in \rf{28} by taking 
the size of all particles equal to the unity. In this way the 
problem is transformed into the evaluation of the eigenvalues 
of a regular (all particles with size 1) inhomogeneous transfer 
matrix $T_0$ with $n(2N-1)$ non-zero vertex and twisted boundary 
condition. 

\noindent {\bf Diagonalization of $T_0(k)$} \\
The simplest way to diagonalize $T_0$ is through the introduction of the 
monodromy matrix ${\cal M} (k)$ \cite{takh}, which is a transfer matrix of the 
inhomogeneous vertex model under consideration, where the first 
 and last link in the horizontal direction are 
fixed to the values $\mu_1$ and $\mu_{n+1}$ ($\mu_1, \mu_{n+1} =1,2,\ldots,N$), that is 
\bea \label{I1}
{\cal M}_{\{Q'\},\mu_1}^{\{Q\},\mu_{n+1}}(k) &=& \Phi_{\mu_1}
\sum_{\mu_2,\ldots,\mu_n=1}^N 
\tilde S_{Q_1',\mu_1}^{Q_1,\mu_2}(k_{P_1},k)
\tilde S_{Q_2',\mu_2}^{Q_2,\mu_3}(k_{P_2},k) \cdots \nonumber \\
&&\cdots \tilde S_{Q_{n-1}',\mu_{n-1}}^{Q_{n-1},\mu_n}(k_{P_{n-1}},k)
\tilde S_{Q_{n}',\mu_{n}}^{Q_{n},\mu_{n+1}}(k_{P_{n}},k).
\eea
The monodromy matrix ${\cal M}_{\{Q'\},\mu_1}^{\{Q\},\mu_{n+1}}(k)$  
 has coordinates $\{Q\},\{Q'\}$ in the vertical 
space ($N^n$ dimensions) and coordinates $\mu_1,\mu_{n+1}$ in the 
horizontal space ($N^2$ dimensions).  
This matrix satisfy the following important relations 
\bea \label{I2}
\sum _{\nu'_1,\mu'_1=1}^N\tilde S_{\nu_1,\mu_1}^{\nu_1',\mu_1'}(k',k)
{\cal M}_{\{\alpha_l\},\mu_1'}^{\{\gamma_l\},\mu_{n+1}}(k)
{\cal M}_{\{\gamma_l\},\nu_1'}^{\{\beta_l\},\nu_{n+1}}(k') &=& \nonumber \\
\sum_{\nu'_{n+1},\mu'_{n+1}=1}^N
{\cal M}_{\{\alpha_l\},\nu_1}^{\{\gamma_l\},\nu_{n+1}'}(k')
&&{\cal M}_{\{\gamma_l\},\mu_1}^{\{\beta_l\},\mu_{n+1}'}(k)  
\tilde S_{\nu_{n+1}',\mu_{n+1}'}^{\nu_{n+1},\mu_{n+1}}(k',k),
\eea
for $\mu_1,\nu_1,\mu_{n+1},\nu_{n+1}=1,2,\ldots,N$.This relation 
 follows 
directly from successive applications of the Yang-Baxter equations \rf{36} 
(see \cite{revbra1}, for a graphical representation of these equations).

In order to exploit relation \rf{I2} let us denote the components of the  
monodromy matrix in the horizontal space by
\bea \label{I3}
A(k)_{\alpha}^{\beta}& =&  {\cal M}_{\{\alpha_l\},\alpha}^
{\{\gamma_l\},\beta}(k), \quad 
B(k)^{\alpha} =  {\cal M}_{\{\alpha_l\},N}^
{\{\gamma_l\},\alpha}(k), \nonumber \\
C(k)^{\alpha}& =&  {\cal M}_{\{\alpha_l\},\alpha}^
{\{\gamma_l\},N}(k), \quad 
D(k) =  {\cal M}_{\{\alpha_l\},N}^
{\{\gamma_l\},N}(k),
\eea
where ($\alpha,\beta =1,2,\ldots,N-1$).
 Clearly the transfer matrix $T_0(k)$ of the  inhomogeneous 
lattice with twisted boundary conditions, we want  to diagonalize, 
is given by 
\beq \label{I4}
T_0(k) = \sum_{\alpha=1}^{N-1}A_{\alpha}^{\alpha}(k) + D(k).
\eeq
As a consequence of \rf{I2} the matrices $A_{\alpha}^{\alpha}$, 
$B^{\alpha}$, $C^{\alpha}$ and $D$ in \rf{I3} 
obey some algebraic relations. By setting $(\nu_1,\mu_1,\nu_{n+1},\mu_{n+1}) 
= (N,\alpha,\gamma,\beta)$ in \rf{I2} we obtain
\beq \label{I5}
A_{\alpha}^{\beta}(k)B^{\gamma}(k') = 
-\frac{\tilde S_{N,\alpha}^{\alpha,N}(k',k)}
{\tilde S_{N,\alpha}^{N,\alpha}(k',k)}
B^{\beta}(k)A_{\alpha}^{\gamma}(k') +
 \sum_{\alpha',\beta'=1}^{N-1}
\frac{\tilde S_{\alpha',\beta'}^{\gamma,\beta}(k',k)}
{\tilde S_{N,\alpha}^{N,\alpha}(k',k)}
B^{\alpha'}(k')A_{\alpha}^{\beta'}(k), 
\eeq
with ($\alpha,\beta=1,\ldots,N-1$). By setting $(\nu_1,\mu_1,\nu_{n+1},
\mu_{n+1}) = (N,N,N,\alpha)$ we obtain
\beq \label{I6}
D(k)B^{\alpha}(k') = \frac{\tilde S_{N,N}^{N,N}(k,k')}
{\tilde S_{N,\alpha}^{N,\alpha}(k,k')}B^{\alpha}(k')D(k) -
 \frac{\tilde S_{\alpha,N}^{N,\alpha}(k,k')}
{\tilde S_{N,\alpha}^{N,\alpha}(k,k')}B^{\alpha}(k)D(k'), 
\eeq
where ($\alpha=1,\ldots,N-1$).
 The diagonalization of $T_0(k)$ in \rf{I4} will be done by
exploiting the above relations. This procedure is known in the literature 
as the algebraic Bethe ansatz \cite{takh}. The first step in this method follows 
from the identification of a reference state $|\Omega>$, which should be  
an eigenstate of $A_{\alpha}^{\alpha}(k)$ and $D(k)$, and hence $T_0(k)$, 
but not of $B^{\alpha}(k)$.  
In the present case a suitable  reference state is 
$|\Omega> = |\{\alpha_l=N\}>_{l=1,\ldots,n}$, which corresponds to a state 
with N-class particles only. It is simple to calculate 
\bea \label{I7}
A_{\alpha}^{\beta}(k)|\Omega> &=& a_{\alpha}^{\beta}(k)|\Omega >, \quad \quad 
D(k)|\Omega> =d(k)|\Omega >, \nonumber \\
C^{\alpha}(k)|\Omega> &=& 0, \quad \quad 
B^{\alpha}(k)|\Omega> =\sum_{i=1}^nb_i^{\alpha}(k)|\Omega_{\alpha}^{(i)} >, 
\eea
where 
\bea \label{I8}
a_{\alpha}^{\beta}(k) &=& \delta_{\alpha,\beta} \Phi_{\alpha} 
\prod_{i=1}^n \tilde S_{N,\alpha}^{N,\alpha}(k_{P_i},k), \quad \quad 
d(k) = \Phi_N \prod_{i=1}^n \tilde S_{N,N}^{N,N}(k_{P_i},k), \quad \quad \nonumber \\
b_i^{\alpha}(k) &=& \Phi_N 
\prod_{l=1}^{i-1} \tilde S_{N,N}^{N,N}(k_{P_l},k)
\prod_{l=i}^n \tilde S_{N,\alpha}^{N,\alpha}(k_{P_l},k),  
\eea
and  $|\Omega_{\alpha}^{(i)} > = |\{\alpha_{l \neq i} =N\},
\alpha_i =\alpha >$.
The matrices  $B^{\alpha}(k)$ act as  creation operators   
in the reference (``vacuum") state, by creating particles of 
class $\alpha$ ($1,2,\ldots,N$) 
 in a sea of particles of Nth class $|\Omega>$. We then 
expect that the  eigenvectors of $T_0(k)$ corresponding 
to $m_1$ ($1, 2, \ldots,n$) 
 particles, belonging to classes distinct from $N$, 
can be expressed as 
\beq \label{I9}
|k_l^{(1)};F> = \sum_{\{\beta\}} F_{\beta_1,\ldots,\beta_{m_1}} 
B^{\beta_1}(k_1^{(1)})B^{\beta_2}(k_2^{(1)})\cdots 
B^{\beta_{m_1}}(k_{m_1}^{(1)})|\Omega >,
\eeq
where $\{k_l^{(1)}, l=1,\ldots,m_1\}$ and 
$F_{\beta_1,\ldots,\beta_{m_1}}$ are variables to be fixed by the 
eigenvalue equation
\beq \label{I10}
T_0(k) |k_l^{(1)},F> = 
\Lambda^{(0)}(k)|k_l^{(1)},F>.
\eeq

Using \rf{I5} successively, and \rf{I7},\rf{I8} we obtain
\bea \label{I11}
A_{\alpha}^{\alpha}(k)B^{\beta_1}(k_1^{(1)})B^{\beta_2}(k_2^{(1)}) 
\cdots B^{\beta_{m_1}}(k_{m_1}^{(1)}) = 
\sum_{\{\alpha_1',\ldots,\alpha_{m_1}'=1\}}^N
\sum_{\{\beta_1',\ldots,\beta_{m_1}'=1\}}^N&&\nonumber \\
\tilde S_{\alpha_1',\beta_1'}^{\beta_1,\alpha} (k_1^{(1)},k) 
\tilde S_{\alpha_2',\beta_2'}^{\beta_2,\beta_1'} (k_2^{(1)},k) \cdots 
\tilde S_{\alpha_{m_1 -1}',\beta_{m_1-1}'}^
{\beta_{m_1-1},\beta_{m_1-2}'} (k_{m_1-1}^{(1)},k) &&
\tilde S_{\alpha_{m_1}',\alpha}^
{\beta_{m_1},\beta_{m_1-1}'} (k_{m_1}^{(1)},k)  \nonumber \\
\times \Phi_{\alpha}
\frac{\prod_{j=1}^n \tilde S_{N,\alpha}^{N,\alpha}(k_j,k) }
{\prod_{j=1}^{m_1} \tilde S_{N,\alpha}^{N,\alpha}(k_j^{(1)},k) }
B^{\alpha_1'}(k_1^{(1)})B^{\alpha_2'}(k_2^{(1)}) 
\cdots B^{\alpha_{m_1}'}(k_{m_1}^{(1)}) |\Omega >  &+& 
{\mbox {"unwanted terms"}},
\eea
where the "unwanted terms" are those ones which are not expressed in the 
"Bethe basis" produced by the $B^{\alpha}(k_j^{(1)})$ operators.
Similarly, using \rf{I6} successively and \rf{I7}-\rf{I8} we obtain
\bea \label{I12}
D(k)B^{\beta_1}(k_1^{(1)}) \cdots B^{\beta_{m_1}}(k_{m_1}^{(1)} )
|\Omega>  =  
 \Phi_N \left(\prod_{l=1}^n 
 \tilde S_{N,N}^{N,N} (k_l,k)\right) && \nonumber \\ 
\times \; \left( \prod_{l=1}^{m_1} 
\frac{\tilde S_{N,N}^{N,N} (k, k_l^{(1)})}
{\tilde S_{N,\beta_l}^{N,\beta_l}(k,k_l^{(1)}) }\right)  
B^{\beta_1}(k_1^{(1)})\cdots B^{\beta_{m_1}}(k_{m_1}^{(1)})
|\Omega> &+&
{\mbox {"unwanted terms"}}.
\eea
The relations \rf{I11} and \rf{I12} when used in \rf{I9}-\rf{I10} 
give us 
\bea \label{I13}
T_0(k)&&|k_{\alpha}^{(1)},F> = 
\frac{\prod _{j=1}^n \tilde S_{N,1}^{N,1}(k_j,k)}
{\prod _{j=1}^{m_1} \tilde S_{N,1}^{N,1}(k_j^{(1)},k)}
\sum_{\{\beta\}} \sum_{\{\alpha'\}}  
T_1(k)_{\{\alpha' \}}^{\{\beta\}} F_{\{\beta\}}  
B^{\alpha'_1}(k_1^{(1)})\cdots B^{\alpha'_{m_1}} (k_{m_1}^{(1)}) 
|\Omega>   \nonumber \\
&&+  
\Phi_N 
\prod_{i=1}^n \tilde S_{N,N}^{N,N} (k_i,k)  
\sum_{\{\beta\}} 
\prod_{l=1}^{m_1}  
\left(
\frac 
{\tilde S_{N,N}^{N,N}(k,k_l^{(1)})}
{\tilde S_{N,\beta_l}^{N,\beta_l}(k,k_l^{(1)})}
\right)
 F_{\{\beta\}} 
B^{\beta_1}(k_1^{(1)})\cdots B^{\beta_{m_1}}(k_{m_1}^{(1)}) 
|\Omega>   \nonumber  \\
&&+ \mbox {"unwanted terms"},
\eea
where
\beq \label{I14}
T_1(k)_{ \{\alpha'\}}^{\{\beta\}}  = 
\sum_{\alpha=1}^{N-1}\Phi_{\alpha} 
\sum_{\beta_1',\ldots,\beta_{m_1}'=1}^{N-1} 
\tilde S_{\alpha_1',\beta_1'}^{\beta_1,\alpha}(k_1^{(1)},k) 
\left( \prod_{i=1}^{m_1-1} 
\tilde S_{\alpha_i',\beta_{i+1}'}^{\beta_i,\beta_i'}
(k_i^{(1)},k) \right) 
\tilde S_{\alpha_{m_1}',\alpha}^{\beta_{m_1},\beta_{m_1 -1}'}
(k_m^{(1)},k)
\eeq
is a $(N-1)^{m_1}$-dimensional transfer matrix of a inhomogeneous 
vertex model, with inhomogeneties 
$\{k_{m_1}^{(1)}, k_{m_1-1}^{(1)}, \ldots,k_1^{(1)} \}$ 
(notice the reverse order of the inohomogeneties, 
when compared with \rf{b2a}) and 
twisted boundary conditions (boundary phases 
$\Phi_{\alpha}, \alpha =1, \ldots,N-1$). 

In order to proceed we need now to diagonalize the new transfer 
matrix $T_1(k)$, that is we must solve 
\beq \label{I15}
\sum_{\{\beta\}} T_1(k)_{ \{\alpha'\}}^{\{\beta\}} 
F_{\{\beta\}} = 
\Lambda^{(1)}(k) F_{\{\alpha'\}}
\eeq
and then \rf{I13} give us
\beq \label{I16}
T_0(k) |k_{\alpha}^{(1)},F> = 
\Lambda^{(0)}(k)|k_{\alpha}^{(1)},F> + \mbox{ "unwanted terms"},
\eeq
where, using the fact that $\tilde S_{N,N}^{N,N}(k_l,k)=1$,
\beq \label{I17}
\Lambda^{(0)}(k) = \frac
{\prod_{j=1}^n \tilde S_{N,\alpha}^{N,\alpha}(k_j,k)}
{\prod_{j=1}^{m_1} \tilde S_{N,1}^{N,1}(k_j^{(1)},k)}
\Lambda^{(1)}(k) + \Phi_N 
\prod_{l=1}^{m_1} \frac{1}
{\tilde S_{N,1}^{N,1}(k,k_l^{(1)})}.
\eeq
In order to proof that $\Lambda^{(0)}$ and 
$|K_{\alpha}^{(1)},F>$ are the eigenvalues and eigenvectors 
of $T_0(k)$, we should fix $\{k_1^{(1)},\ldots,k_{m_1}^{(1)}\}$
 by requiring that the "unwanted terms" in \rf{I16} vanish. 
Although for $N=2$ this calculation is not complicated 
\cite{revbra1} for arbitrary $N$ it is not simple. 
Since the  expression \rf{I17} for the eigenvalues should be 
valid for arbitrary values of $k$ we can obtain 
$\Lambda^{(1)}(k_j^{(1)})$ in an alternative way 
from the following trick 
\cite{kulish}. At $k=k_j^{(1)}$ ($j=1,\ldots,m_1$) 
the denominators of the factors in \rf{I17} vanishes 
($\tilde S_{n,l}^{N,l}(k_j^{(1)},k_j^{(1)}) =0 
, l\neq N$), and since we should have a finite result, 
we have the conditions
\beq \label{I18}
\Lambda^{(1)}(k_j^{(1)}) = \Phi_N 
\prod_{i=1}^n \frac{1} {\tilde S_{N,1}^{N,1}(k_i,k_j^{(1)})}
\prod_{l'=1,l'\neq j}^{m_1} \frac 
{\tilde S_{N,1}^{N,1}(k_{l'}^{(1)},k_j^{(1)})}
{\tilde S_{N,1}^{N,1}(k_{j}^{(1)},k_{l'}^{(1)})},\, j=1,\ldots,m_1.
\eeq
Notice that our result in \rf{I18} does not depend on the 
particular ordering of the additional variables 
$k_j^{(1)}$ ($j=1,\ldots,m_1$). This means that if instead of 
the ordering chosen in \rf{I9}, we chose the reverse order, namely,
\beq \label{I19}
|k_{\alpha}^{(1)},F> = \sum_{\{\beta\}} 
F_{\beta_1,\ldots,\beta_{m_1}} 
B^{\beta_{m_1}}(k_{m_1}^{(1)})
B^{\beta_{m_1 -1}}(k_{m_1-1}^{(1)}) \cdots
B^1(k_1^{(1)}) |\Omega>
\eeq
we obtain the same results \rf{I16}-\rf{I18} but now $T_1$ is 
the transfer matrix, with boundary condition specified by 
the phase $\Phi_{\alpha}$, of a problem with $(N-1)$ species 
and inhomogeneities 
$k_1^{(1)},\ldots,k_{m_1}^{(1)}$ (notice we have now the same order
 of the inhomogeneties as in \rf{b2a}). 
This means that the eigenvalue 
$\Lambda(k)=\Lambda^{(0)}(k)$ of the transfer matrix of the 
problem with $N$ classes and inhomogeneities 
($k_1^{(0)},k_2^{(0)},\ldots,k_n^{(0)}) \equiv 
(k_1,k_2,\ldots,k_n$) is 
related to the eigenvalue $\Lambda^{(1)}(k)$ of the problem 
with ($N-1$) classes  and inhomogeneities $k_1^{(1)},k_2^{(1)},\ldots,
k_{m_1}^{(1)}$. Iterating these calculations we obtain the 
generalization of the relation \rf{I17} and the 
condition \rf{I18}
\bea \label{I20}
\Lambda^{(l)}(k) &=& 
\left( \prod_{l'=1}^{m_l} \tilde S_{N,1}^{N,1}(k_{l'}^{(l)},k)
\right) 
\left( \prod_{l'=1}^{m_{l+1}}  \frac
{1} {\tilde S_{N,1}^{N,1}(k_{l'}^{(l+1)},k)} \right)
\Lambda^{(l+1)}(k) +  \nonumber \\
&&\Phi_{N-l} \prod_{l'=1}^{m_{l+1}} \frac
{1} {\tilde S_{N,1}^{N,1}(k,k_{l'}^{(l+1)})}, 
l =0,1,\ldots, N-1,
\eea
\beq \label{I21}
\Lambda^{(l+1)}(k_j^{(l+1)}) = \Phi_{N-l} 
\prod_{l'=1}^{m_l} \left( \frac {1} 
{\tilde S_{N,1}^{N,1}(k_{l'}^{(l)},k_j^{(l+1)})} \right)
\prod_{\l'=1,l'\neq j}^{m_{l+1}} \frac 
{\tilde S_{N,1}^{N,1}(k_{l'}^{(l+1)},k_j^{(l+1)})}
{\tilde S_{N,1}^{N,1}(k_{j}^{(l+1)},k_{l'}^{(l+1)})},
\eeq
which connects the eigenvalues of the inhomogeneous 
transfer matrix $T_l(k)$ and $T_{l+1}(k)$, with inhomogeneities 
$\{k_j^{(l)}\}$ and $\{k_j^{(l+1)}\}$, related with the 
problem with $(N-l)$ and $(N-l-1)$ classes of particles, 
respectively.

However from \rf{40} and \rf{b2}-\rf{b2a}, in order to obtain 
the Bethe-ansatz equations for our original problem we 
need the eigenvalues of the transfer matrices evaluated 
at $k_j$ ($j=1,\ldots,n$), i. e., $\Lambda^{(0)}(k_j)$, 
which are given by
\beq \label{I22}
\Lambda^{(0)}(k_{P_j}) = \Phi_N \prod_{l=1}^{m_1} 
\frac{1} {\tilde S_{N,1}^{N,1}(k_{P_j},k_l^{(1)})},
\eeq
since $\prod_{j=1}^n \tilde S_{n,\alpha}^
{N,\alpha}(k_j,k_{P_j}) =0$.
The conditions that fix the variables 
($k_j^{(1)}, j=1,\ldots,m_1$) are given by \rf{I18}. 
In the left side of this equation we have 
$\Lambda^{(1)}(k_j^{(1)})$, which are the eigenvalues 
of the transfer matrix $T_1$ of the model with $(N-1)$ 
classes of particles and inhomogeneities $\{k_j^{(1)}, 
j=1,\ldots,m_1\}$, evaluated at the partcular point 
$k_j^{(1)}$. This value can be obtained from \rf{I20} 
 which gives a generalization of \rf{I22} 
\beq \label{I23}
\Lambda^{(l)}(k_j^{(l)}) =  \Phi_{N-l} 
\prod_{l'=1}^{m_{l+1}} \frac {1} 
{\tilde S_{N,1}^{N,1}(k_j^{(l)},k_{l'}^{(l+1)})} 
\quad (l =0,1,\ldots,N-1).
\eeq
The condition \rf{I18} is then replaced by 
\beq \label{I24}
\Lambda^{(1)}(k_j^{(1)}) = \Phi_{N-1} 
\prod_{l'=1}^{m_2} \frac{1} 
{\tilde S_{N,1}^{N,1}(k_j^{(1)},k_{l'}^{(2)})} = 
 \Phi_{N} 
\prod_{i=1}^{n} \frac{1} 
{\tilde S_{N,1}^{N,1}(k_i^{(1)},k_{j}^{(1)})} 
\prod_{l'=1,l'\neq j}^{m_1} \frac 
{\tilde S_{N,1}^{N,1}(k_{l'}^{(1)},k_j^{(1)})} 
{\tilde S_{N,1}^{N,1}(k_{j}^{(1)},k_{l'}^{(1)})} ,
\eeq 
where now we need to find the relations that fix 
$\{k_j^{(2)}\}$. Iterating this process we find the 
generalization of \rf{I24}
\bea \label{I25}
 \Phi_{N-l} && \prod_{l=1}^{m_{l+1}} \frac{1}
{\tilde S_{N,1}^{N,1}(k_j^{(l)},k_{l'}^{(l+1)})} =  
 \Phi_{N-(l-1)} \prod_{i=1}^{m_{l-1}}  \frac {1}
{\tilde S_{N,1}^{N,1}(k_{i}^{(l+1)},k_j^{(l)})} 
\nonumber \\
&& \times \prod_{l'=1,l' \neq j}^{m_l} \frac
{\tilde S_{N,1}^{N,1}(k_{l'}^{(l)},k_j^{(l)})}
{\tilde S_{N,1}^{N,1}(k_{j}^{(l)},k_{l'}^{(l)})}, 
(j=1,2,\ldots,m_l;\quad l=0,1,\ldots,N-2).
\eea
Equations  \rf{I22} and \rf{I25} give us the eigenvalues 
of the tranfer matrix $T_0(k)$ evaluated at the points 
$\{k_j\}$, i. e. $\Lambda^{(0)}(k_j)$. Inserting the 
above results in \rf{b2} and then in \rf{40} we obtain 
the Bethe-ansatz equations of our original problem. 

The 
eigenenergies of the Hamiltonian \rf{6} in the sector containing 
$n_i$ particles in class $i$ ($i=1,2,\dots,N$) ($ n = \sum_{j=1}^Nn_j$) 
and total momentum $p=\frac{2\pi}{L}$ ($l=0,1,\dots,L-1$) are 
given by 
\beq \label{64}
E = -\sum_{j=1}^n(\epsilon_-e^{ik_j^{(0)}} +
\epsilon_+e^{-ik_j^{(0)}} -1),
\eeq
where $\{k_j^{(0)} = k_j, j =1,\dots,n\}$ are obtained from the 
solutions $\{k_j^{(l)}, l=0,\dots,N-1; j=1,\dots,m_l\}$ of the Bethe 
ansatz equations
\bea \label{65}
e^{ik_j(L+n-\sum_{i=1}^N n_is_i)} &=& (-1)^{n-1} e^{-ip(s_N-1)} 
\prod_{j'=1 (j' \neq j)}^n 
\frac{\epsilon_+ + \epsilon_-e^{i(k_j^{(0)}+k_{j'}^{(0)})} -
e^{ik_j^{(0)}}}
{\epsilon_+ + \epsilon_-e^{i(k_j^{(0)}+k_{j'}^{(0)})} -
e^{ik_{j'}^{(0)}}} \nonumber \\
&\times & \prod_{l=1}^{m_1} 
\frac{\epsilon_+(e^{ik_l^{(1)}} - e^{ik_j^{(0)}})}
{\epsilon_+ + \epsilon_-e^{i(k_l^{(1)}+k_{j}^{(0)}} -
e^{ik_{j}^{(0)}}} \quad j=1,2,\dots,n,
\eea
and
\bea \label{66}
&& \prod_{\beta=1}^{m_l} \frac{\epsilon_+(e^{ik_{\alpha}^{(l+1)}} 
- e^{ik_{\beta}^{(l)}})}
{\epsilon_+ + \epsilon_-e^{i(k_{\alpha}^{(l+1)}+k_{\beta}^{(l)})} -
e^{ik_{\beta}^{(l)}}}   = (-1)^{m_{l+1}} e^{ip(s_{N-l}-s_{N-l-1})} 
 \prod_{\delta=1}^{m_{l+2}} \frac{\epsilon_+(e^{ik_{\delta}^{(l+2)}} 
- e^{ik_{\alpha}^{(l+1)}})}
{\epsilon_+ + \epsilon_-e^{i(k_{\delta}^{(l+2)}+k_{\alpha}^{(l+1)})} -
e^{ik_{\alpha}^{(l+1)}}}   
\nonumber \\
&&\times \prod_{\alpha'=1 (\alpha' \neq \alpha)}^{m_{l+1}} 
\frac{\epsilon_+ + \epsilon_-e^{i(k_{\alpha}^{(l+1)}+k_{\alpha'}^{(l+1)})} -
e^{ik_{\alpha}^{(l+1)}}}
{\epsilon_+ + \epsilon_-e^{i(k_{\alpha}^{(l+1)}+k_{\alpha'}^{(l+1)})} -
e^{ik_{\alpha'}^{(l+1)}}} \quad l=0,1,\dots,N-2;\quad \alpha =1,\dots,m_l,
\eea
and $m_l = \sum_{j=1}^{N-l}n_j$, $l=0,\ldots,N$ ($m_0=N,m_N=0$).
It is interesting to observe that in the particular case where 
$n_2=n_3 =\dots=n_N=0$ we 
obtain the Bethe-ansatz equations, recently derived \cite{alc-bar1} 
(see also \cite{sasawada}),  
 for the 
asymmetric diffusion problem with particles of size $s_1$. Also the case  
$s_1=s_2=\dots =s_N =1$ 
give us the corresponding Bethe ansatz equations for the  
standard problem of $N$ type of particles in hierarchical order. 
The Bethe-ansatz solution in the particular case of N=2 with a 
a single particle of class 2 ($n_1=n-1,n_2=1$) was derived recently 
\cite{deri2}.
The Bethe-ansatz equations for 
the fully asymmetric problem is obtained by setting 
in \rf{65}-\rf{66} $\epsilon_+=1$ and
 $\epsilon_- =0$.

\section{ Conclusions and generalizations }

We obtained through the Bethe ansatz the exact solution of the problem  
 where particles belonging to $N$ distinct classes with  hierarchical 
order diffuse as well interchange positions with rates depending on their 
relative hierarchy. We show that the exact solution can also be derived 
in the general case where the particles have arbitrary sizes. 

Some extensions of our results can be made. A first  and quite 
interesting  generalization of our model 
 happens when we  allow 
 molecules in any class to have size  $s=0$. Molecules 
of size zero do not occupy space on the lattice, having no hard-core 
exclusion effect. Consequently we may have, at a given lattice point,  
an arbitrary number of them. 
The Bethe-ansatz solution presented in the previous 
section is extended directly in this case (the equations are the same) 
and the eigenenergies are given 
by fixing in \rf{64}-\rf{66} the appropriate sizes of the molecules. It 
is interesting to remark that particles of a given  class $c'$ 
($2,3,\dots,N$), with size $s_{c'}=0$,
contrary to the case  $s_{c'}>1$, where they ``accelerate" 
the diffusion of the 
 particles in classes $c <c'$, 
now they ``retard" the diffusive motion of these 
particles. The quantum Hamiltonian in the cases where the particles have  
size zero  is obviously not given by 
\rf{6} but can be  written in terms of spin $S=\infty$ quantum chains.
Another further extension of our model is obtained by considering an 
arbitrary mixture of molecules, where  molecules in the same 
hierarchy may have distinct sizes. The results presented in 
\cite{alc-bar1} correspond to the particular case of this generalization 
where  $N=1$ (simple 
diffusion). For general $N$ the $S$ matrix we obtain in \rf{28} is 
also a solution of the Yang-Baxter equation \rf{36}, but the 
diagonalization of the transfer matrix of the associated 
inhomogeneous vertex model is more complicated.
The Bethe-ansatz equations in the case of the asymmetric 
diffusion, with particles of unit size \cite{spohn,kim}, or with 
arbitrary size \cite{alc-bar1}, 
were used to obtain the finite-size corrections of the 
mass gap $G_N$ of the associated quantum chain. The real part of these 
finite-size corrections are governed by the dynamical critical exponent 
$z$, i. e., 
\beq
\mbox {Re}(G_N) \sim N^{-z}. \nonumber
\eeq
The calculation of the exponent $z$ for the model presented in this paper, 
with particles of arbitrary sizes, is presently in progress \cite{alc-bar2}.

\acknowledgements {
This work was supported in part by Conselho Nacional de Desenvolvimento
Cient\'{\i}fico e Tecnol\'ogico - CNPq - Brazil and
by the Russian Foundation of Fundamental Investigation ( Grant 99-02-17646). }

\newpage 
%


\newpage
\Large
\begin{center}
Figure Captions
\end{center}
\normalsize
\vspace{1cm}

\noindent Figure 1 - Example of configurations of molecules with distinct 
sizes $s$ in a lattice of size $L=6$. The coordinates of the molecules  
are denoted by the black squares. 

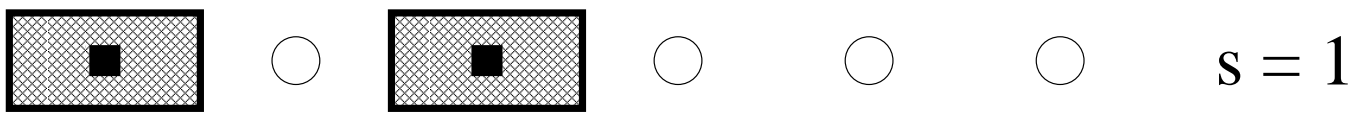
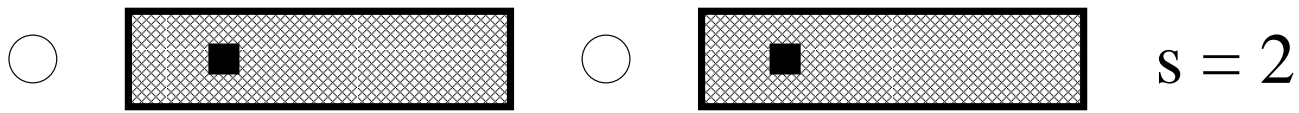
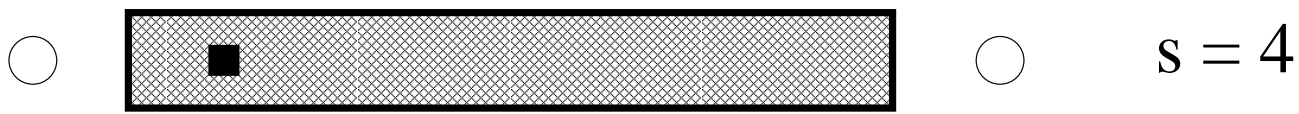
\end{document}